\documentclass[a4paper,11pt]{article}
\usepackage{aaskaiid}
\usepackage{xcolor}
\usepackage{needspace}
\usepackage{orcidlink}
\usepackage{caption}

\title{A comprehensive SKA Survey of lensing clusters and lensed galaxies}
\ShortTitle{Gravitational Lensing with clusters of galaxies}

\author[1]{M. Pandey-Pommier\orcidlink{https://orcid.org/0000-0001-5829-1099}}
\ShortName{M. Pandey-Pommier et al.} 
\author[2]{J. Mckean \orcidlink{https://orcid.org/0000-0003-1787-9552}}
\author[3]{I. Harrison \orcidlink{https://orcid.org/0000-0002-4437-0770}}
\author[4]{C. Cress \orcidlink{https://orcid.org/0000-0003-3483-3951}}

\affiliation[1]{Pole Scientific, University Catholic of Lyon- University of Lyon, 10 place des Archives 69288, Lyon, France}
\emailAdd{mamtapommier@gmail.com}

\affiliation[2]{University of Pretoria, Private Bag x20 Hatfield 0028, South Africa}
\emailAdd{john.mckean@up.ac.za}

\affiliation[3]{School of Physics and Astronomy, Cardiff University, CF24 3AA, United Kingdom}
\emailAdd{harrisoni@cardiff.ac.uk}

\affiliation[4]{Mathematical Sciences, University of South Africa, Christiaan de Wet Rd, Florida, 1709, Johannesburg, South Africa}
\emailAdd{cresscm@unisa.ac.za}

\abstract{Galaxy clusters act as powerful gravitational lenses that provide a unique laboratory for studying the distribution of dark matter and the properties of faint background galaxies. In the radio regime, cluster lensing studies remain largely unexplored due to the intrinsically low surface density of background radio sources, which limits the number of detectable lensed systems and constrains the application of cluster-scale lensing to cosmological and astrophysical investigations. In this paper, we outline how the SKA  will transform cluster lensing studies through its unprecedented sensitivity, angular resolution, and survey speed. These capabilities will enable dense sampling of lensed background radio sources and high-fidelity reconstruction of cluster mass distributions, while revealing a previously inaccessible population of faint ($\mu$Jy-level) galaxies, primarily at $z \sim 1-5$. We focus on three key science goals: (i) constraining dark matter substructure in galaxy clusters through improved strong-lensing constraints, (ii) characterising the non-thermal baryonic components of the intracluster medium, and (iii) building statistically significant samples of $\mu$Jy-level lensed galaxies to study star formation and active galactic nuclei activity across cosmic time (primarily at $z \sim 1-5$).  Together, these advances will establish cluster-scale radio lensing as a powerful probe of structure formation and galaxy evolution in the SKA era.} 

\begin{document}
\maketitle

\section{Introduction}
Gravitational lensing is an astronomical phenomenon where the gravitational pull of a dense cosmic object, such as a cluster or galaxy, bends the light from a distant background galaxy. This bending of light gives rise to the curvature of spacetime, resulting in multiple images of the distant galaxy \citep{Zwicky1937}.
Depending on the alignment of the distant galaxy and mass distribution of the lensing system, lensing can manifest in various forms, including strong and weak lensing \citep{Kochanek2004}. This phenomenon can produce a range of effects, including distorted images, large arc-like or multiple symmetric structures, or Einstein rings, all of which align along the critical curves of the lensing system \citep{schneider2006gravitational, Virbhadra1998}. 

Galaxy clusters represent the most massive gravitationally bound systems in the Universe and are among the most powerful gravitational lenses known due to their enormous mass (up to 10$^{15}M_\odot$), a critical requirement for cosmic lensing. While $\sim 70-80\%$  of the total mass in massive galaxy clusters is dominated by dark matter, followed by $\sim 15-20\%$ of hot gas (T $\sim10^{15}$K) in the Intra Cluster Medium (ICM), the remaining is in the form of a few percent of baryonic matter in the galaxies. The dark matter dominates the cluster potential and is expected to be collision less in nature, interacting through gravity only \citep{Bradac2006, Natarajan2024}. Cold Dark Matter predicts significant sub-structure in halos, and lensing provides one of the few probes of this prediction  \citep{Powell2025}. Through strong and weak lensing, clusters provide direct constraints on the projected total mass distribution, enabling detailed studies of dark matter structure and its interaction with baryonic components. Strong gravitational lensing in clusters enables us to probe the high-redshift galaxies, including those from the epoch of reionization $z \sim6$ and beyond, that would otherwise be too faint to observe. The magnified lensed images offer critical insights into galaxy evolution and star formation in the distant universe \citep{schneider2006gravitational, Umetsu2016, Ueda2019}. In addition, lensing analyses provide key insights into the internal structure of galaxy clusters, including their substructures and mass distributions. Combined with X-ray and radio observations, these analyses offer a more comprehensive view of the baryonic and dark matter components of clusters, as well as their complex interplay \citep{Zitrin2015, PandeyPommier2016, Natarajan2024}.

This work is motivated by three fundamental questions:
\begin{itemize}
\item  What is the distribution and level of substructure in the dark matter halos of lensing clusters?  
\item  How do baryonic processes such as hot gas dynamics and non-thermal plasma relate to the underlying dark matter potential?  
\item  What is the nature of faint radio-emitting galaxies at intermediate and high redshift, and how are they magnified by cluster lenses?
\end{itemize}

To address these questions, cluster lensing observations require both high-resolution mass mapping and a sufficiently dense background population of detectable sources. Extensive multi-wavelength observational campaigns such as the ESO Hamburg survey \citep{Wisotzki1996}, CfA-Arizona-ST-LEns-Survey (CASTLES by \citep{Munoz1998}; Frontier Fields (FF) program using the HST \citep{lotz2017ApJ...837...97L}, ALMA \citep{Hezaveh2013}, JWST \citep{Castellano2023}, and MUSE/VLT \citep{Richard2015} have significantly deepened our understanding of lensing clusters' structure, mass profiles, substructure, and their role in magnifying faint, high-redshift background galaxies. In addition, the SDSS survey uncovered hundreds of lensed galaxies and quasars \citep{Bolton2008, Inada2012} while wide-field surveys such as DES \citep{DESCollaboration2016}, and Euclid \citep{Laureijs2011} have discovered several new gravitational lenses and enabled statistical studies of cluster mass distributions. Collectively, these surveys have expanded the catalog of strong lenses at redshifts as high as, $z \sim1.98$ and allowed detailed reconstructions of lensed galaxies out to, $z \approx 10$ \citep{Negrello2010, Vieira2013, Finner2025}.

At radio wavelengths, galaxy-galaxy gravitational lensing has been extensively explored through surveys such as the Jodrell Bank–VLA Astrometric Survey (JVAS) and the Cosmic Lens All Sky Survey (CLASS) \citep{Browne1998, Browne2003, Patnaik1992, Myers1995, Myers2003, Jackson1995, McKean2015}. 
These surveys used flat-spectrum radio sources to discover the majority of known radio gravitational lenses. To date, approximately $300$-$500$ lensing systems have been identified across different mass scales (galaxies and groups), though only about $\sim10\%$ involve radio-loud background sources.
These surveys have enabled lensing investigations up to redshifts of $z \sim 1.5$ \citep{Jackson1995, McKean2007}, although they typically rely on ancillary optical or infrared data for redshift determination.  In addition, recent studies on lensing cluster-galaxy systems with the SKA pathfinders, GMRT, and LOFAR confirmed the dynamical state of the lensing clusters \citep{PandeyPommier2016} and at higher frequencies ($\sim 1-6$ GHz) and resolution ($< 1.8$) arcsec with the JVLA discovered several $(\sim 20)$ radio-emitting lensed galaxies up to $(z \sim  2)$ \citep{vanWeeren2016, Heywood2021}.  However, these cluster-scale lensing studies in the radio regime remain vastly under-utilised, especially limited to a couple of FF clusters up to $(z \sim  0.545)$.  In general, current radio facilities are still limited by sensitivity and angular resolution, resulting in a small number of detected lensed sources and insufficient background source density for detailed cluster mass reconstruction \citep{McKean2015, PandeyPommier2018, Heywood2021}. These limitations reflect both selection biases and instrumental constraints, as present-day radio surveys are optimised for compact lens configurations and are not yet sensitive enough to routinely detect the $\mu$Jy-level background source population required for robust cluster-scale lensing analyses. Further, while gravitational lensing robustly constrains the total projected mass distribution, it does not directly probe gas dynamics. Instead, multi-wavelength observations combining lensing (mass), X-ray emission (hot gas), and radio emission (non-thermal plasma) are required to disentangle the baryonic and dark matter components of clusters.  Furthermore, at 
$(> z \sim  0.54)$, our understanding of mass distributions and the dark–baryonic matter interplay in lensing clusters still heavily depends on optical, infrared, and mm-observations, as radio observations at these redshifts remain largely unexplored \citep{Natarajan2024, Finner2025}. 


In this paper, we focus on cluster-galaxy lensing as a probe of dark matter substructure and the population of faint lensed radio sources. We investigate how increased source density and improved angular resolution of SKA will enable more accurate mass reconstruction of cluster lenses and open the possibility of building statistically meaningful samples of  $\mu$Jy-level lensed galaxies at $z \sim 1-5$, providing new constraints on galaxy evolution in dense environments. Throughout the paper, we adopt a $\mathrm{\Lambda CDM}$ cosmology with $\mathrm{H_{0}=70}$
km $\mathrm{s^{-1}Mpc^{-1}}$, $\Omega_{M} = 0.3$ and $\Omega_{\Lambda} =0.7$.

\needspace{3\baselineskip}
\section{Baryonic and Dark matter interplay in Lensing clusters}
\subsection{Tracing Matter Distribution via Low-Frequency Radio Emission in Lensing Clusters}
Low-frequency radio observations are crucial for understanding the distribution of matter, non-thermal component of the ICM, and the dynamical state in lensing galaxy clusters. The degree of merger activity in a cluster’s central region determines the type of non-thermal radio emission observed in the ICM, which is typically faint and more effectively detected at low radio frequencies due to its steep spectral nature ($\alpha \lesssim -1.3$).  It is important to note here that these radio-emitting structures are not unique to lensing-selected clusters, but instead are a general property of massive and dynamically active galaxy clusters. Lensing clusters are therefore best understood as a biased subset of the most massive and centrally concentrated clusters, which are particularly efficient laboratories for studying both gravitational lensing and diffuse radio emission due to their high total mass and strong gravitational potential.

Depending on their physical origin and location within the broader class of massive cluster, the non-thermal radio emissions are classified into the strongly polarized peripheral `relics' (a few kpc - Mpc scale) and central fossil plasma `Phoenix' (a few kpc scale), both of which arise from shocks or compression within the ICM. 
Additionally, turbulence induced by mergers or gas sloshing generates diffuse, unpolarized centrally located radio structures, such as, haloes ($>$ 500 kpc, up to or beyond 1 Mpc) in dynamically unrelaxed, elongated `non-cool core' (NC) clusters, and more compact mini-haloes ($<$ 500 kpc) in relaxed, compact `cool core' (CC) clusters \citep{Feretti2012}. In addition, relaxed CC clusters tend to 
host a central radio-loud bright cluster galaxy (BCG), while non-relaxed NCC clusters often contain multiple radio-emitting ellipticals in their center associated with the sub-group undergoing merging activity. 

Over the past decade, SKA pathfinder surveys such as uGMRT, JVLA, LOFAR and MeerKAT,  have significantly expanded the number of known diffuse radio sources in clusters, including rare ultra-steep spectrum radio halos (USSRHs), thanks to their $\mu$ Jy-level sensitivity down to 110 MHz. USSRHs are characterised by very steep spectral indices ($\alpha \lesssim -1.5$), making them preferentially detectable at low radio frequencies. They are thought to trace less energetic cluster mergers and therefore provide important probes of particle acceleration mechanisms and the dynamical evolution of galaxy clusters \citep{Venturi2008, Cassano2010, Bonafede2012, kale2013,kale2015, PandeyPommier2013, Brunetti2014, PandeyPommier2016, vanweeren2019, Cuciti2021, Knowles2022, Phuravhathu2025}. They therefore provide key constraints on particle acceleration efficiency and the dynamical state of clusters that may otherwise appear relaxed at higher frequencies. However, the observed abundance of these systems remains lower than theoretical expectations, highlighting a discrepancy between predicted and observed populations of diffuse radio emission \citep{Cassano2015}. The first-ever targeted low-frequency radio survey of lensing clusters (up to 
$z \sim 0.545$) was conducted by \citep{PandeyPommier2016},  finding relics, halos, and mini-halos in those systems, similar to those in non-lensing clusters. However, compared to non-lensing clusters, lensing clusters tend to be massive (often $\ge 0.5 \times 10^{15}  M_{\odot}$) and generally show high central mass concentrations and relaxed CC nature (with few exceptions from merging clusters, e.g., MACS J$0717.5+3745$) (ref. Fig.1).
\begin{figure}[t]
    \centering
    \includegraphics[width=0.51\textwidth]{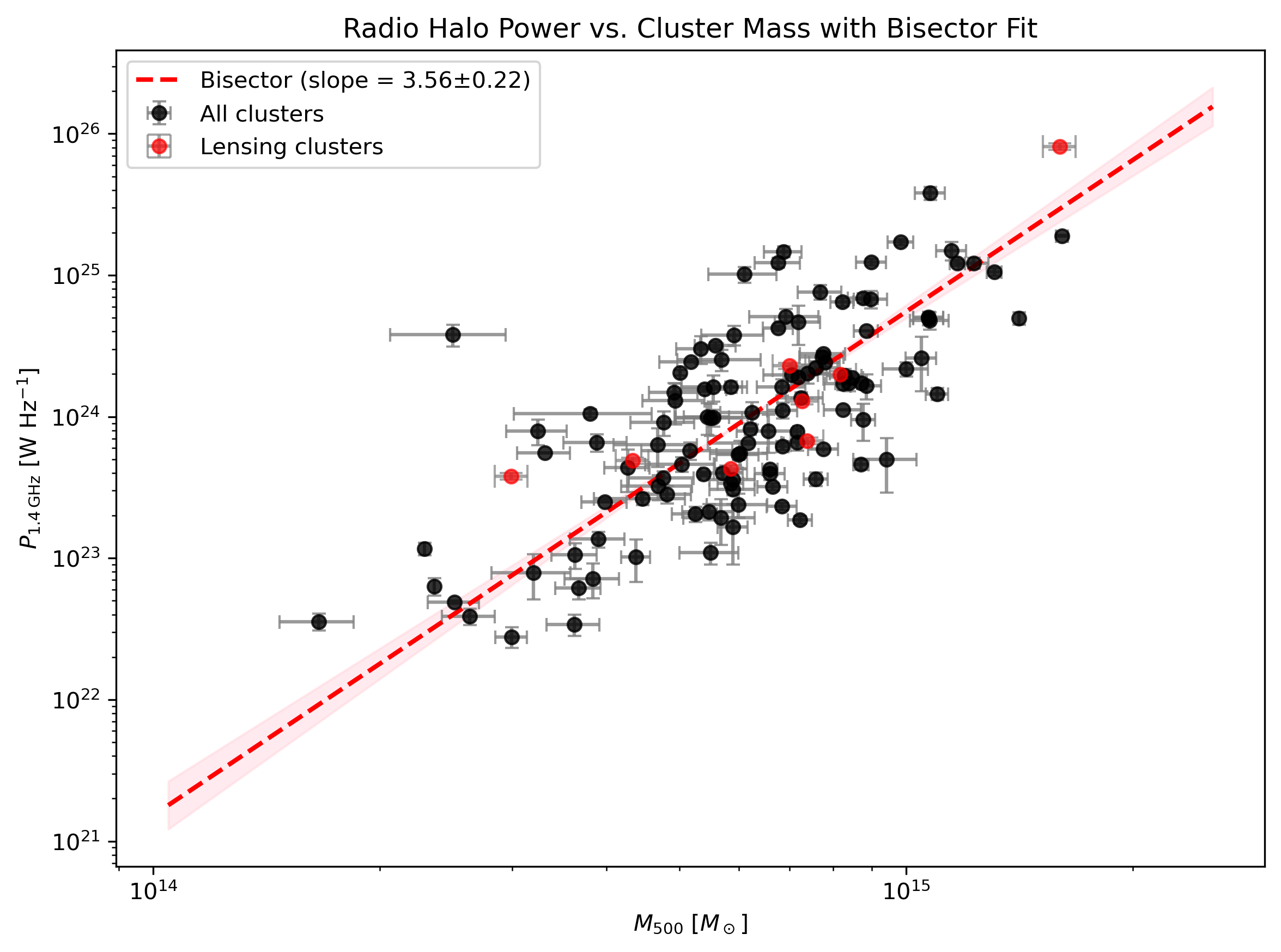}
    \includegraphics[width=0.48\textwidth]{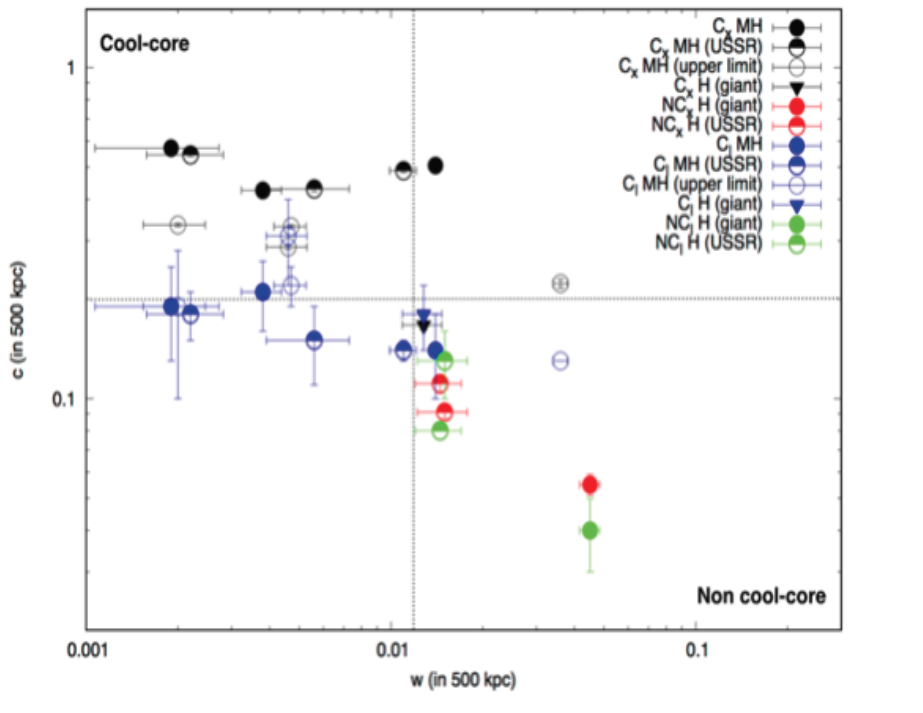}
   \caption{{\em Left panel}: Radio power-mass distribution  
   diagram for the clusters with known radio halos from literature for lensing (red) and non-lensing (black) clusters. {\em Right panel}: Dark-baryonic matter coupling in lensing galaxy clusters. The non-detection limits are marked as `UL', radio halo as `RH', X-ray concentration in cool-core as `Cx', X-ray concentration in non cool-core as `NCx', concentration in total mass map via lensing analysis as `Cl' in cool-cores and `NCl' in non cool-core clusters \citep{PandeyPommier2016}.
   }
    \label{fig:radiopower}
\end{figure}
The discrepancies in radio halo number counts is largely driven by current limitations in sensitivity and surface brightness detection, which the SKA will address by enabling a statistically complete census of diffuse radio emission in galaxy clusters. 

\subsection{Gas$-$Mass Alignment and the Dark–Baryonic Interplay in Lensing Galaxy Clusters}
Gravitational lensing provides a direct and robust reconstruction of the projected total mass distribution in galaxy clusters, making it one of the most powerful probes of the underlying dark matter potential and of the spatial relationship between dark and baryonic matter. It is important to note that lensing clusters are not representative of the general cluster population; rather, they constitute a biased subset toward high masses and centrally concentrated mass distributions, which enhance their lensing cross-sections, increase their detectability in lensing surveys, and amplify observable signatures of baryonic$-$dark matter interplay.

Multiwavelength observations combining gravitational lensing (total mass), X-ray emission (hot ICM), and radio emission (non-thermal plasma) are essential to fully characterise the dynamical state of these systems. While lensing and X-ray data provide constraints on the collisional (gas) and collisionless (dark matter) components, radio observations offer an additional and independent tracer of non-thermal processes driven by cluster dynamics. In particular, diffuse radio emission such as halos, mini-halos, and relics traces turbulence, merger-driven shocks, and particle acceleration within the ICM, while spectral index variations provide information on the energy distribution and ageing of relativistic electrons. In addition, radio emission associated with AGN in BCGs probes feedback processes that regulate the thermodynamic state of the cluster core. These radio observables therefore provide complementary and physically independent constraints on cluster dynamical activity that are not accessible through lensing or X-ray data alone. Morphological diagnostics derived from X-ray and lensing maps, such as the concentration parameter ($c$) and centroid shift ($w$), are widely used to quantify cluster relaxation states \citep{Cassano2010, kale2015, PandeyPommier2016}. As illustrated in Fig.~1 (left panel) lensing clusters occupy the high-mass end of the radio power–mass relation. This reflects the selection bias toward massive and centrally concentrated systems with enhanced lensing cross-sections.  In Fig.~1 (right panel), relaxed CC systems exhibit centrally concentrated X-ray emission and well-aligned mass and gas peaks, indicating a close spatial correspondence between baryonic and dark matter distributions. In contrast, non-relaxed NCC systems show disturbed morphologies, with significant offsets between the X-ray peak and the lensing-derived mass centroid. These offsets arise from merger-induced shocks and bulk gas motions, which affect the collisional baryonic component, leaving the collision-less dark matter distribution largely unaffected.  The morphology, and spectral properties of these radio components provide a complementary diagnostic of cluster assembly and dynamical activity, complementary to X-ray and lensing-based measures.
%

Owing to their large masses and strong gravitational potentials, lensing clusters provide a unique opportunity to study the coupling between dark matter, baryons, and non-thermal plasma, enabling direct comparison with baryonic tracers across wavelengths. Current progress, however, is constrained by limited sensitivity and angular resolution for detecting faint diffuse radio emission and mapping its connection to cluster mass substructures.


\subsection{Probing Cold Gas Reservoirs in Lensing Galaxy Clusters}
Relaxed CC clusters are characterised by a central concentration of cold molecular gas, which provides a striking contrast to the hotter X-ray emitting ICM in the outer regions \citep{Fabian1994, Edge2001}. These regions are strongly influenced by feedback from the central AGN, which regulates the thermodynamic state of the ICM through a balance between radiative cooling and mechanical heating from jets and outflows. In low-entropy systems, reduced cooling times allow gas to condense out of the hot phase, forming molecular reservoirs that can fuel both star formation and AGN activity. This cooling$-$feedback cycle is therefore directly linked to the baryonic evolution of the central BCGs and may correlate with lensing efficiencies through their association with dynamically relaxed, centrally concentrated systems. Recent observations by \citep{Castignani2020} investigated molecular gas in 18 BCGs belonging to the CLASH lensing cluster sample over the redshift range $z \approx 0.2-0.9$, using CO line observations (Fig.~2). Only one system (RXJ1532.9+3021) shows a robust detection of a substantial molecular gas reservoir, while four additional BCGs exhibit tentative detections. The remaining 13 systems show very low molecular gas-to-stellar mass ratios ($M_{\mathrm{H2}}/M_\star < 0.1$), indicating that most BCGs in this sample are relatively poor in cold molecular gas. The detected molecular gas masses span $M_{\mathrm{H2}} \sim 10^{10}-10^{11} \, M_\odot$, with RXJ1532.9+3021 exhibiting a particularly large reservoir of $M_{\mathrm{H2}} = (8.7 \pm 1.1)\times10^{10} \, M_\odot$ and an associated star formation rate (SFR) exceeding $100 \, M_\odot \, \mathrm{yr}^{-1}$. 
\begin{figure}[h]
    \centering
    \includegraphics[width=0.97\textwidth]{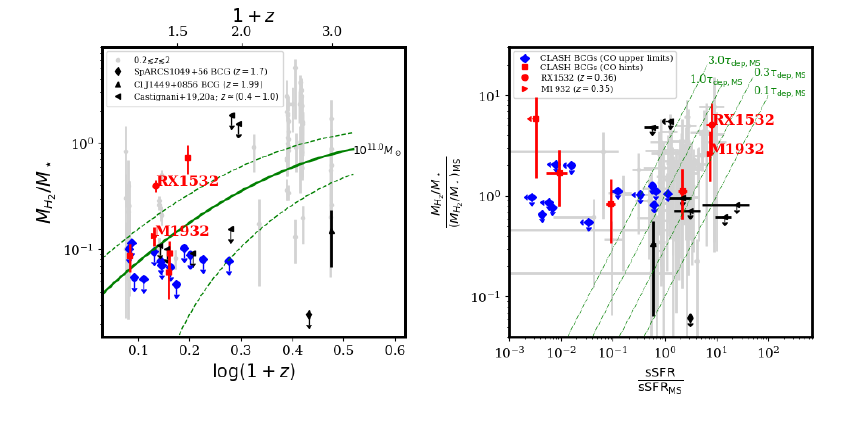}
    \caption{Molecular gas properties of distant BCGs in lensing clusters observed in CO ({\em Left panel}) with IRAM and their star formation rate ({\em Right panel}), from \citep{Castignani2020}}
    \label{fig:lensing_schematic}
\end{figure}
The left panel of Fig.~2 shows a weak but important redshift dependence with higher-redshift systems tend to have less constrained molecular gas measurements due to observational sensitivity limits, rather than intrinsic absence of gas. This highlights the need for larger and more sensitive samples, particularly at higher redshift, to determine whether the observed trends reflect true evolution or selection effects. These findings are consistent with the picture that most gas-rich BCGs are preferentially located in cool-core systems, as suggested by previous studies \citep{Edge2001}. BCGs with high SFRs tend to host significant $H_{2}$ masses and display clumpy, filamentary morphologies, underscoring the role of central entropy and cooling flows in driving molecular gas accumulation. At the same time, mechanical feedback from the central AGN, through radio jets and gas sloshing, plays a key role in regulating gas inflow and suppressing cooling. BCGs with low central entropy frequently exhibit both emission-line signatures of ongoing star formation \citep{Fogarty2015} and evidence of AGN activity \citep{Hogan2015}, suggesting a common fuelling source linked to the hot ICM and a close interplay between cooling flows and AGN feedback. In relaxed cool-core systems, this AGN feedback cycle can explain the coexistence of molecular gas, star formation, and AGN activity in a subset of low-entropy clusters, while efficient feedback heating keeps many other systems relatively gas poor.

Radio observations provide an independent probe of this feedback cycle. In the CLASH sample, \citep{Heng2018} find that 1.5 GHz radio power in BCGs spans $P_{1.5\,\mathrm{GHz}} \sim 2\times10^{23}$ to $10^{26}$ W Hz$^{-1}$, with a statistically significant positive correlation with SFR and a negative correlation with ICM entropy.These results indicate that stronger AGN activity is preferentially associated with systems where cooling is more effective, consistent with a self-regulated feedback scenario.  From a lensing perspective, these baryonic processes are particularly relevant because low-entropy, dynamically active cores often correspond to high central mass concentrations and enhanced lensing efficiency.  However, current observational constraints remain incomplete as: (i) the sample size of molecular gas measurements in lensing-selected BCGs is small, (ii) the redshift coverage is limited, particularly beyond $z \gtrsim 0.5$, and (iii) sensitivity limitations prevent detection of low-mass gas reservoirs in typical systems. In addition, most current surveys lack uniform frequency coverage for simultaneously tracing both molecular gas and atomic gas in the same cluster environments. The SKA will overcome these limitations by enabling significantly deeper observations.

\needspace{3\baselineskip}
\subsection{Background Lensed Galaxies}
Background-lensed galaxies in the environments of massive lensing clusters detected as unresolved radio sources or arcs, offer valuable insights into galaxy formation and evolution at high redshifts, \citep{PandeyPommier2018}. Using deep JVLA observations of the massive FF lensing cluster MACS J0717.5+3745, \citep{vanWeeren2016} reported the detection of several strongly lensed background radio sources out to $z\sim $2.1. In a complementary study, \citep{Heywood2021} conducted high-resolution radio imaging on MACS J0717.5+3745 and MACS J1149.5+2223, at 3 GHz and 6 GHz using the JVLA. These observations achieved sub-arcsecond resolution and $\mu$Jy-level sensitivity, reaching RMS noise levels of ~5–10 $\mu$Jy beam$^{-1}$, enabling detection of both cluster-associated and gravitationally lensed background galaxies out to $z\sim $1.2. The detected lensed sources exhibit flux densities typically ranging from a few $\mu$Jy to 100s of $\mu$Jy, with many detections near the 10-30 $\mu$Jy level, barely at the detection limit of JVLA. After correcting for lensing magnification, these correspond to intrinsically low-luminosity radio sources in the $\mu$Jy-level regime. However, a discrepancy in the redshift distribution of detected lensed sources was observed between studies, primarily due to \citep{Heywood2021} adopting median magnification values, whereas \citep{vanWeeren2016} used mean values.
\begin{figure}[b]
    \centering
    \includegraphics[width=0.41\textwidth]{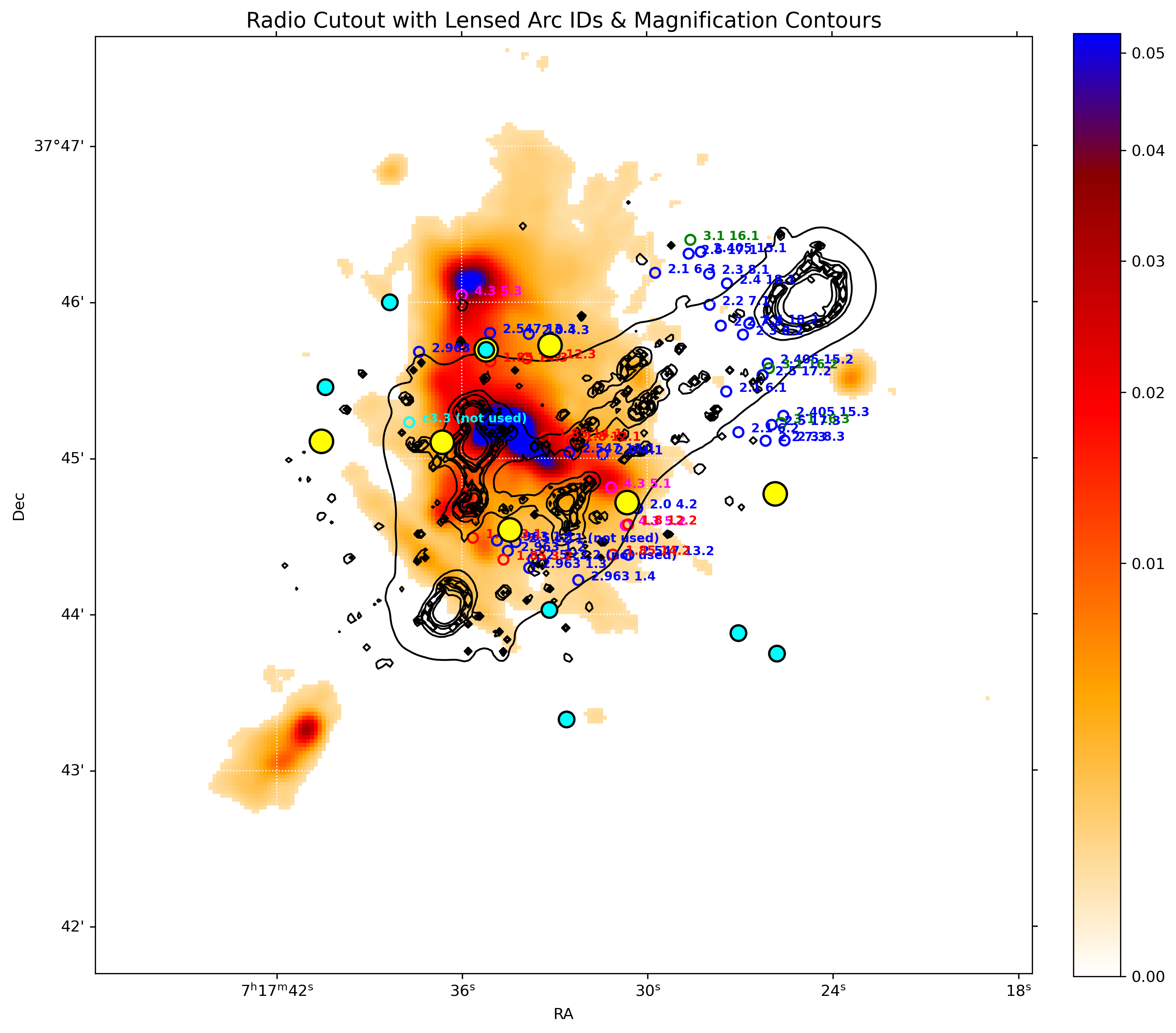}
        \includegraphics[width=0.58\textwidth]{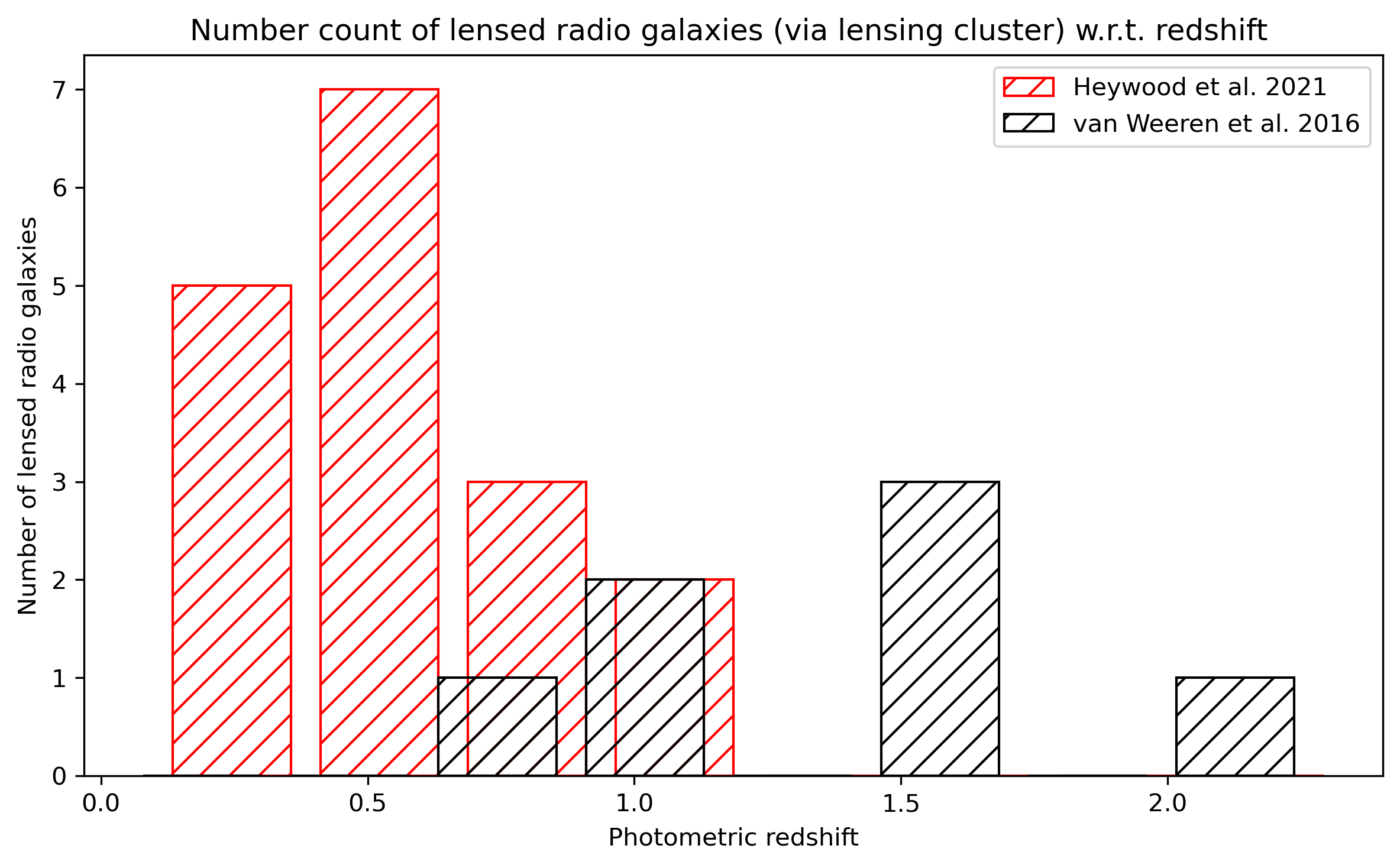}
    \caption{{\em Left panel}: Gravitational Lensing in MACSJ0717.5+3745 cluster with total mass map in black contours overlaid on GMRT data \citep{PandeyPommier2016} showing JVLA radio lensed galaxies in cyan (\cite{vanWeeren2016}) and yellow (\citep{Heywood2021}) circles. {\em Right panel}: Radio lensed galaxies discovered in massive clusters environments.}
    \label{fig:lensing_schematic}. 
\end{figure}
Based on radio morphology, spectral properties, and multiwavelength counterparts, the lensed radio galaxies observed in the FF fall into star-forming galaxies, typically late-type or irregular systems, exhibiting compact, unresolved radio emission consistent with non-thermal synchrotron radiation. Their optical and near-infrared counterparts often show blue continuum and clumpy, irregular morphologies, particularly at redshifts  $z >$1. In addition, a subset of sources is associated with red, compact galaxies, likely quiescent ellipticals hosting low-luminosity AGN that show weak radio emission and little to no star formation activity, possibly indicating radio-mode AGN feedback. A few galaxies also showed an intermediate nature where both AGN and star formation features were observed simultaneously. 
Such sources may represent transition phases in galaxy evolution, possibly linked to episodes of AGN triggering, feedback-regulated star formation, or merger-driven gas inflows in dense environments. However, current observations remain limited to determine whether these hybrid sources correspond to a distinct evolutionary stage or arise from multiple physical mechanisms. 

\section{SKA capabilities and multi-wavelength synergies}
As discussed in previous sections, despite several advances, current radio lensing studies still face two major observational challenges: (1) insufficient sensitivity to detect large numbers of faint background sources, and (2) limited angular resolution for resolving multiple lensed images in cluster environments. As a result, constraints on cluster substructure and high-redshift lensed galaxy populations remain incomplete in the radio band. With $\mu$Jy-level sensitivity, sub-arcsecond resolution, and wide-field survey capability,  the SKA will increase the number density of detectable background radio sources by orders of magnitude, enabling high-fidelity reconstruction of cluster mass distributions and the identification of statistically significant samples of strongly lensed galaxies \citep{Prandoni2014, McKean2015, PandeyPommier2016, PandeyPommier2018}. These advances will establish cluster-scale radio lensing as a powerful probe of dark matter substructure, baryonic processes, and galaxy evolution through gravitational magnification.

\begin{figure}[b]
   \centering
   \includegraphics[width=0.9\textwidth]{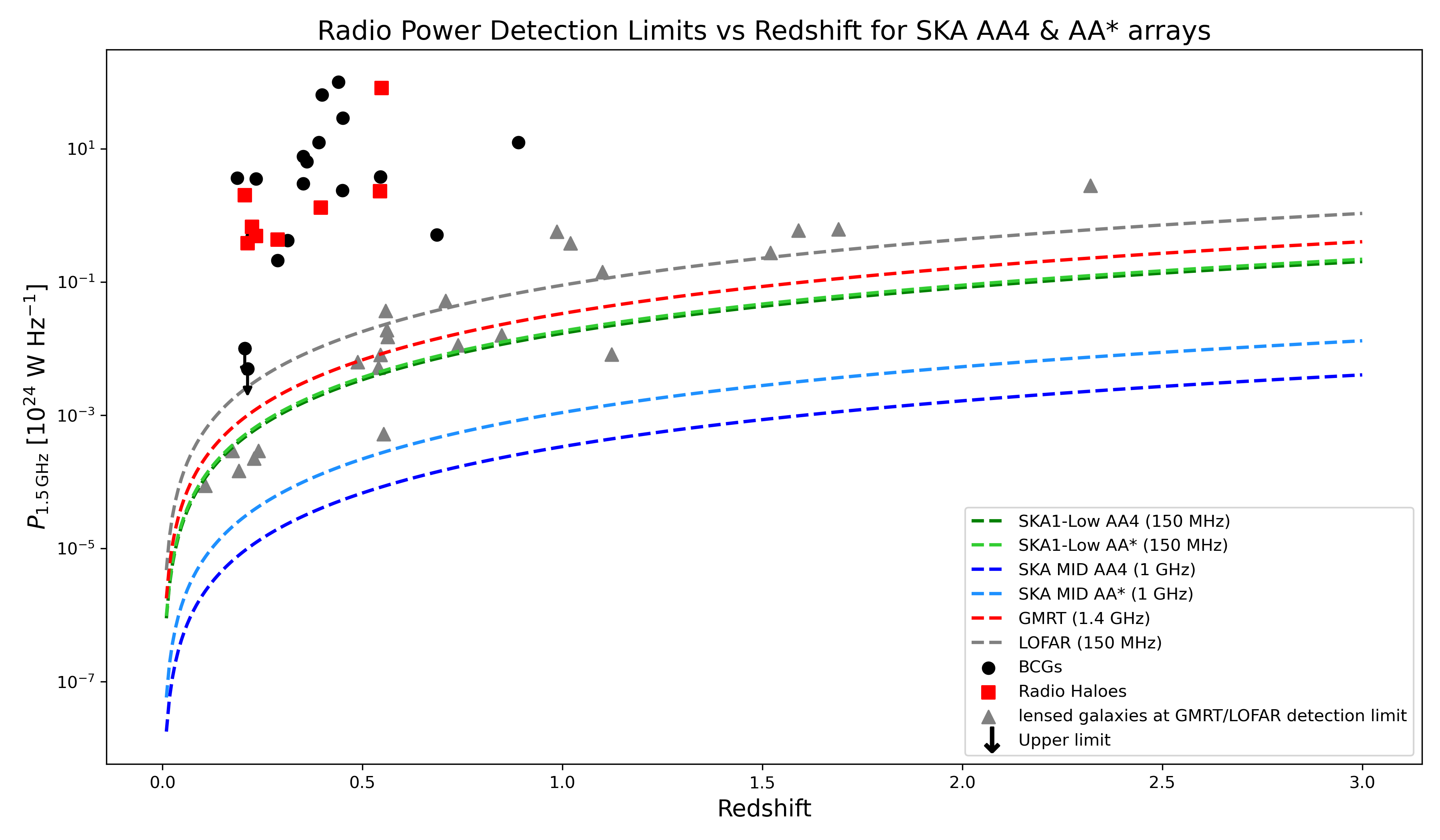}
    \caption{
Radio array sensitivities in comparison to the SKA AA4 and AA* arrays, showing the radio power at 1\,GHz detectable with a sensitivity of 1\,$\mu$Jy (defined as 3$\times$ the RMS noise), as computed using the SKA sensitivity calculator: \href{https://sensitivity-calculator.skao.int}{sensitivity-calculator.skao.int}. 
The black, red, and grey points indicate radio luminosities currently measured in BCGs, radio haloes observed with GMRT and LOFAR, and lensed distant galaxies with the JVLA, respectively, in lensing clusters~\citep{Heng2018, PandeyPommier2016, vanWeeren2016, Heywood2021}.}
\label{fig: Instrument sensitivities}
\end{figure}

As shown in Fig.~4, SKA AA* and AA4 arrays and current path-finders are already capable of detecting diffuse synchrotron emission from cluster haloes and relics in massive systems, while higher-frequency instruments such as the JVLA provide the angular resolution required to identify compact lensed radio galaxies. However, existing observations remain sensitivity limited, particularly for faint $\mu$Jy-level background galaxies and high-redshift diffuse radio emission. 
SKA-LOW and SKA-MID will overcome these limitations by combining high surface-brightness sensitivity with sub-arcsecond resolution, enabling simultaneous studies of cluster radio haloes, relics, and strongly lensed galaxies across a broad redshift range. This will allow direct investigation of the interplay between cluster dynamics, dark matter structure, AGN feedback, and galaxy evolution across cosmic time  \citep{McKean2015, PandeyPommier2018}. A major advance enabled by SKA  (AA* and AA4) arrays will be the detection of large populations of faint lensed radio galaxies at $z\sim1-5$, many of which are currently inaccessible. These observations will constrain the faint end of the radio luminosity function and provide new insights into the co-evolution of star formation and AGN activity in low-luminosity galaxies. In addition, high-resolution radio imaging of multiply imaged systems will improve constraints on cluster substructure and the distribution of dark matter within cluster cores.

Beyond continuum observations, SKA spectral-line observations will probe the cold gas content and kinematics of cluster galaxies through H~{\sc i} 21-cm transition and low-$J$ CO emission, enabling measurements of gas reservoirs, inflows/outflows, and star-formation fueling in cluster galaxies. These data will be key to understanding cooling flows, AGN feedback, and star-formation regulation in cluster cores \citep{Pandey-Pommier03.2026.SKA}. At higher frequencies, SKA Band 5 may also detect dense gas tracers such as HCN and HCO$^+$ in highly magnified, strongly lensed systems at $z \gtrsim 5$, offering rare constraints on dense star-forming gas in the early Universe \citep{Pandey-Pommier01.2026.SKA} . Combined with lensing magnification, this will enable detailed studies of the interstellar medium in faint, high-redshift galaxies near the peak of cosmic star formation and potentially into the epoch of reionization.

The scientific return of SKA observations will be significantly enhanced through multi-wavelength synergies with forthcoming facilities. Optical/near-IR surveys from Euclid, LSST, JWST and WST \citep{Mainieri2024} will provide photometric and spectroscopic redshifts, stellar masses, and detailed morphological information for lensed galaxies and cluster members  \citep{Laureijs2011, Castellano2023}. ATHENA (X$-$rays) will trace the thermodynamic state of the hot ICM, while ALMA/NOEMA will probe the molecular gas component through CO imaging and spectroscopy  \citep{Barcons2017, Wootten2009}. Together, these datasets will deliver a comprehensive multi-phase view of baryons in clusters, linking total mass distributions from lensing to hot, cold, and non-thermal components. Further, in combination with optical weak-lensing surveys, SKA data will further improve cluster mass reconstructions and tighten constraints on structure formation and the evolution of massive halos across cosmic time.



\section{Summary and conclusion}
In this chapter, we have shown that galaxy clusters are powerful gravitational lenses and key laboratories for studying the interplay between dark matter, baryons, and non-thermal processes through radio observations. While current data already reveal diffuse emission, gas$-$mass offsets, and a limited population of lensed radio galaxies, progress is constrained by sensitivity and resolution limits. The SKA will overcome these challenges with $\mu$Jy sensitivity and high angular resolution, enabling detailed studies of cluster substructure, diffuse ICM emission, and large samples of strongly lensed galaxies beyond $z\sim2$.  Additionally, the spectral-line capabilities of SKA will provide direct constraints on the cold gas content and kinematics of cluster galaxies.  The SKA Band~5 may also enable detections of dense molecular gas tracers (HCN and HCO$^+$) in highly magnified systems and star-forming distant lensed galaxies at  $z \gtrsim 5$.  Together with future multi-wavelength facilities, the SKA will provide a comprehensive view of cluster evolution and lensed galaxy populations across cosmic time.

\section*{Author Ordering}
Authors for this chapter are ordered according to their overall level of expertise and contribution, in line with that expected for a small author list publication.

\section*{Acknowledgements}
We thank the anonymous referee for their constructive comments and insightful suggestions, which helped improve the clarity of this manuscript.

\bibliographystyle{abbrvnat-maxbibnames4}
\bibliography{chapter} 

\end{document}